\documentclass[aps,prb,twocolumn,superscriptaddress,showpacs]{revtex4}

\usepackage{amsmath}
\usepackage{bm}
\usepackage{graphicx}

\usepackage{amsfonts}

\begin{document}

\title{Absence of extended states in a ladder model of DNA}

\author{E.\ D\'{\i}az}
\affiliation{GISC, Departamento de F\'{\i}sica de Materiales, Universidad
Complutense, E-28040 Madrid, Spain}

\author{A.\ Sedrakyan}
\affiliation{Yerevan Physics Institute, Alikhanian Br.\ str.\ 2, 375\,036
Yerevan, Armenia}

\author{D.\ Sedrakyan}
\affiliation{Yerevan Physics Institute, Alikhanian Br.\ str.\ 2, 375\,036
Yerevan, Armenia}

\author{F.\ Dom\'{\i}nguez-Adame}
\affiliation{GISC, Departamento de F\'{\i}sica de Materiales, Universidad
Complutense, E-28040 Madrid, Spain}

\begin{abstract}

We consider a ladder model of DNA for describing carrier transport in a fully
coherent regime through finite segments. A single orbital is associated to each
base, and both interstrand and intrastrand overlaps are considered within the
nearest-neighbor approximation. Conduction through the sugar-phosphate backbone
is neglected. We study analytically and numerically the spatial extend of the
corresponding states by means of the Landauer and Lyapunov exponents. We
conclude that intrinsic-DNA correlations, arising from the natural base pairing,
does not suffice to observe extended states, in contrast to previous claims.

\end{abstract}


\maketitle

\section{Introduction}

\label{intro}

According to standard theories of disordered systems,~\cite{Abrahams79} all
states in low-dimensional systems with uncorrelated disorder are spatially
localized. Therefore, in a pure quantum-mechanical regime, DNA might be
insulator unless the localization length reaches anomalously large values. To
explain long range electronic transport found experimentally,~\cite{Porath00}
Caetano and Schulz claimed that intrinsic DNA-correlations, due to the base
pairing (A--T and C--G), lead to electron delocalization.~\cite{Caetano05}
Furthermore, they pointed out that there is a localization-delocalization
transition (LDT) for certain parameters range. If these results were correct,
then \emph{transverse\/} correlations arising intrinsically in DNA could explain
long range electronic transport. However, we have claimed that this is not the
case and all states remain localized, thus excluding a LDT.~\cite{Sedrakyan06}
Therefore, the electrical conduction of DNA at low temperature is still an open
question.

In this paper we provide further analytical and numerical support to our above
mentioned claim, aiming to understand the role of intrinsic DNA-correlations in
electronic transport. To this end, we address signatures of the  spatial extend
of the electronic states by means of the analysis of the Landauer and Lyapunov
coefficients, to be defined below. The outline of the paper is as follows. In
the next section, we introduce the ladder model of DNA~\cite{Caetano05} and
diagnostic tools we use to elucidate the spatial extend of electronic states in
the static lattice. In Sec.~\ref{Landauer} we discuss the analytical calculation
of the Landauer exponent and show that this exponent never vanishes in the
thermodynamics limit for any value of the system parameters. From this result we
conclude that extended states never arise in the model. We then proceed to
Sec.~\ref{Lyapunov}, in which we numericaly calculate the Lyapunov exponent for
finite samples. We discuss in detail its dependence on the model parameters,
especially inter- and intrastrand hoppings. We provide evidences that the
localization length is only of the order of very few turns of the double helix
for realistic values of the model parameters. Therefore, this shows that
intrinsic DNA-correlations alone cannot explain long range electronic transport.
found in long DNA molecules.~\cite{Porath00}  Finally, Sec.~\ref{conclusions}
concludes the paper.

\section{Model and diagnostic tools} \label{model}

\begin{figure}[ht]
\centering
\includegraphics[width=60mm,clip]{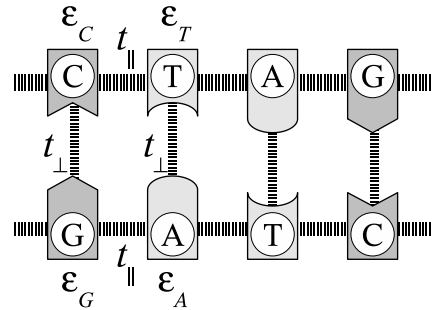}
\caption{Schematic view of a fragment of the ladder model for DNA molecules,
excluding the sugar-phosphate backbone. A single orbital is associated to each
base, with a corresponding energy $\varepsilon_n$, $n$ being A, T,  C or G. The
sequence of the basis of one of the strands is random, while the sequence of the
other strand results from the base pairing A--T and C--G. Only two different
hoppings are considered, namely interstrand ($t_\perp$) and intrastrand
($t_\Vert$) hoppings.}
\label{fig1}
\end{figure}

Our analysis proceeds as follows. We consider a ladder model of DNA in a  fully
coherent regime and assign a single orbital to each base. Conduction through the
sugar-phosphate backbone is neglected hereafter. Both interstrand and
intrastrand overlaps are considered within the nearest-neighbor approximation.
We assume that the hopping does not depend on the base and therefore only two
values are considered, namely interstrand ($t_\perp$) and intrastrand
($t_\Vert$) hoppings. Figure~\ref{fig1} shows a schematic view of a fragment of
this ladder model.

Four different values of the energy sites ($\varepsilon_A$, $\varepsilon_T$,
$\varepsilon_C$, and $\varepsilon_G$) are randomly assigned in one of the
strands, with the same probability, while the sites of the second strand are set
to follow the DNA pairing (A--T and C--G). Hereafter we will restrict ourselves
to the following values of the site energies, taken from
Ref.~\onlinecite{Roche03}, $\varepsilon_{A}=8.24\,$eV,
$\varepsilon_{T}=9.14\,$eV, $\varepsilon_{C}=8.87\,$eV,
$\varepsilon_{G}=7.75\,$eV. The same site energy values were taken in
Ref.~\onlinecite{Caetano05}. As a consequence, only three parameters remain in
the model, namely $t_\perp$, $t_\Vert$ and $N$, the number of base pairs. We
will show below that the spatial extend of the states strongly depends on
$t_\perp$ and $t_\Vert$, but never goes to infinity in the thermodynamics
limit~($N\to\infty$).

Once the model has been established, we can write down the equation for the
amplitudes at different bases. Let us denote these amplitudes as
$\psi_{n}^{(\sigma)}$, where $\sigma=1,2$ runs over the two strands and
$n=1,2,\ldots,N$ denotes the position of the bases at each strand. According to
the model introduced above, the equations for the amplitudes are readily found
to be
\begin{subequations}
\begin{eqnarray}
E\psi_{n}^{(1)}=\varepsilon_{n}^{(1)}\psi_{n}^{(1)}+t_{\Vert}
\left(\psi_{n+1}^{(1)}+\psi_{n-1}^{(1)}\right)+t_\perp\psi_{n}^{(2)}\ ,\\
E\psi_{n}^{(2)}=\varepsilon_{n}^{(2)}\psi_{n}^{(2)}+t_{\Vert}
\left(\psi_{n+1}^{(2)}+\psi_{n-1}^{(2)}\right)+t_\perp\psi_{n}^{(1)}\ .
\label{amplitudes}
\end{eqnarray}
\end{subequations}
Here $\varepsilon_{n}^{(\sigma)}$ takes one of the four values of the site
energies, according to the constraints presented above.

The equation for the amplitudes can be cast in a compact form by using $4\times
4$ transfer matrices. To this end, let us introduce the $4$-vector
\begin{subequations}
\begin{equation}
{\bm\Phi}_n\equiv
\left(\psi_{n}^{(1)},\psi_{n}^{(2)},\psi_{n-1}^{(1)},\psi_{n-1}^{(2)}\right)^t
\label{fourvector}
\end{equation}
where the superscript $t$ indicates the transpose. Defining the following
$2\times 2$ matrix
\begin{equation}
M_{2}^{(n)} \equiv
   \left(
      \begin{array}{cc}
      \frac{E-\varepsilon_{n}^{(1)}}{t_\Vert} & -\,\frac{t_\perp}{t_\Vert}\\
      -\,\frac{t_\perp}{t_\Vert} & \frac{E-\varepsilon_{n}^{(2)}}{t_\Vert}
      \end{array}
   \right)\ ,
\end{equation}
we arrive at the transfer-matrix equation ${\bm\Phi}_{n+1}=T_n{\bm\Phi}_{n}$,
with
\begin{equation}
T_n \equiv
   \left(
      \begin{array}{cc}
      M_{2}^{(n)} & -\mathcal{I}_2\\
      \mathcal{I}_2     & \mathcal{O}_2
      \end{array}
   \right)\ ,
\label{transfermatrix}
\end{equation}
\end{subequations}
where $\mathcal{I}_2$ and $\mathcal{O}_2$ are the unity and null $2\times 2$
matrices, respectively.  One can easily find out, that the $4\times 4$ transfer
matrix $T_n$ satisfies the condition $T_n^\dag J T_n= J$ with
\begin{eqnarray}
\label{J}
J=  \left(
\begin{array}{cccc}
0 & 0 & -i & 0  \\
0 & 0 & 0 & -i \\
i & 0 & 0 & 0 \\
0 & i & 0 & 0
\end{array}
\right)
\end{eqnarray}
which means that $T_n$ belong to the
$SU(2,2)$ group. It is to be noticed that only four transfer matrices appear in
DNA due to the intrinsic pairing (see Fig.~\ref{fig1}), and they will be denoted
as $T_{n,AT}$, $T_{n,TA}$, $T_{n,CG}$, $T_{n,GC}$ for the sake of clarity.

The electronic properties can be described by the full transfer
matrix ${\cal M}_N = \prod_{n=N}T_{n,ij}$ and the Lyapunov
exponents, $\gamma_\textrm{Ly}$, the eigenvalues of the limiting
matrix $\lim_{N\to\infty}\,\ln({\cal M}_N{\cal M}_N^\dag)^{1/2N}$,
provide information about the localization length of the states,
assuming exponential localization.\cite{Kramer93} Here $i,j \in
\{\mathrm{A,T,G,C}\}$ and length is measured in units of the base
separation along a single strand ($3.1\,$nm). Due to the
self-averaging property they can be calculated by taking the product
of random transfer matrices over a long system. Similarly, in the
Landauer exponent $\gamma_\textrm{La}(N)= \ln \langle
\,||\prod_{n=N}^{1} T_{n,ij}||\,\rangle^{1/N}$ (hereafter $\langle
\ldots \rangle$ denotes ensemble averages) is twice the largest
Lyapunov exponent near the critical region in one-dimensional
systems~\cite{Anderson80,SSBS} and can be calculated analytically
following the technique developed in
Refs.~\onlinecite{Sedrakyan99,Hakobyan00,Sedrakyan04}. In quasi-one
dimensional chains, as in the ladder system under consideration,
both exponents again exhibit the same critical behavior (the
critical indices are the same), but their ratio can be different
from 2.

\section{Landauer exponent} \label{Landauer}

In Refs.~\onlinecite{Sedrakyan99, Hakobyan00, Sedrakyan04} it was shown that the
Landauer resistance and the corresponding exponent can be calculated exactly. To
this end, the direct product ${\cal M}_N \otimes {\cal M}_N^\dag$ of the
fundamental representations of transfer matrices of the $SU(1,1)$ group is
reduced to the adjoint one. We apply this technique here for the group
$SU(2,2)$. In order to calculate this direct product exactly, we  use the known
representation of the permutation operator via generators $\tau^\mu$  $ (\mu
=1,\dots,15)$ of the $sl(4) $ algebra as $P=(1/4)(\mathbb{I}\otimes \mathbb{I}+
\tau^{\mu}\otimes\tau^{\mu})$. Thus the matrix elements satisfy
\begin{eqnarray}
\label{Perm}
\delta_{\alpha_1}^{\alpha_2}\delta_{\beta_1}^{\beta_2}=\frac{1}{4}\,
[\delta_{\alpha_1}^{\beta_2}\delta_{\beta_1}^{\alpha_2}+
(\tau^\mu)_{\alpha_1}^{\beta_2}(\tau_\mu)_{\beta_1}^{\alpha_2}]\ ,
\end{eqnarray}
where  we assume summation in the repeated indices $\mu$. Among of generators
$\tau^\mu$ there is one, which coincides with the metric $J$ defined
in~(\ref{J}). We denote the corresponding index $\mu$ as $J$, namely $\tau^J =
J$.

Multiplying~(\ref{Perm}) by $T_j$ and $T^\dag_j$ from the left and right hand
sides respectively, one can express the direct product of $T_j$ and $T^\dag_j$
via their adjoint representation
\begin{eqnarray}
\label{TT}
(T_j)^{\alpha}_{\alpha^{\prime}}(T_{j}^{+})^{\beta^{\prime}}_{\beta}=
{\frac{1}{4}}(J)^{\alpha}_{\beta}
(J)^{\beta^{\prime}}_{\alpha^{\prime}}+ {\frac{1 }{4}}(\tau^{\mu}
J)^{\beta^{\prime}}_{\alpha^{\prime}} \Lambda_{j}^{\mu\nu}(J
\tau^{\nu})^{\alpha}_{\beta}\ .
\end{eqnarray}
Here the adjoint representation $\Lambda_n$  of $T_n$ is defined by
\begin{eqnarray}
\label{Lambda} \Lambda_n^{\mu\nu}=\frac{1}{4}
{\rm Tr}(T_n\tau^{\mu}T^{+}_n\tau^{\nu})\ ,
\end{eqnarray}
being an $15\times 15$ matrix that depends on the parameters of the model at
site $n$ of both chains.

To calculate analytically the Landauer exponent, we apply this decomposition to
the products of fundamental representations, $T_n$'s in ${\cal M}_N$, and after
averaging obtain
\begin{equation}
\langle {\cal M}_N {\cal M}_N^\dag \rangle ={\frac{1 }{4}} J \otimes J +
{\frac{1}{4}} (\tau^{\mu} J)\otimes (J \tau^{\nu})
\left(\prod_{j=1}^{N} \langle\Lambda_{j}\rangle\right)^{\mu \nu}\ .
\label{Gamma}
\end{equation}

It is then straightforward to get the average over four equivalent substitutions
of the base pairs in the random chain
\begin{equation}
\langle \Lambda^{\mu\nu} \rangle = \frac{1}{4}\,\left(
\Lambda^{\mu\nu}_{AT}+\Lambda^{\mu\nu}_{TA}+
\Lambda^{\mu\nu}_{CG}+\Lambda^{\mu\nu}_{GC}\right) \ .
\label{matrix2}
\end{equation}
The Landauer exponents are the nonnegative eigenvalues of $\frac{1}{2}\log\langle \Lambda
\rangle$. The condition of the existence of an extended state is equivalent to
$\det |\langle \Lambda \rangle-I|=0$, and it is a matter of simple algebra to
prove that this condition is never met.

Therefore, we come to the conclusion that the system studied by Caetano and
Schulz~\cite{Caetano05} cannot support truly extended states. Consequently, a
LDT is not to be observed since all states are spatially localized.

\section{Lyapunov exponent} \label{Lyapunov}

It can be argued that, although the localization length is always finite, as we
have demonstrated above, it could be larger than typical sizes used in transport
experiments, as those carried out by Porath \emph{et al.}~\cite{Porath00} To
quantitatively determine the spatial extend of the electronic states, we have
numerically calculated the Lyapunov exponents for different values of the
hoppings $t_\perp$ and $t_\Vert$.

\begin{figure}[ht]
\centering
\includegraphics[width=70mm,clip]{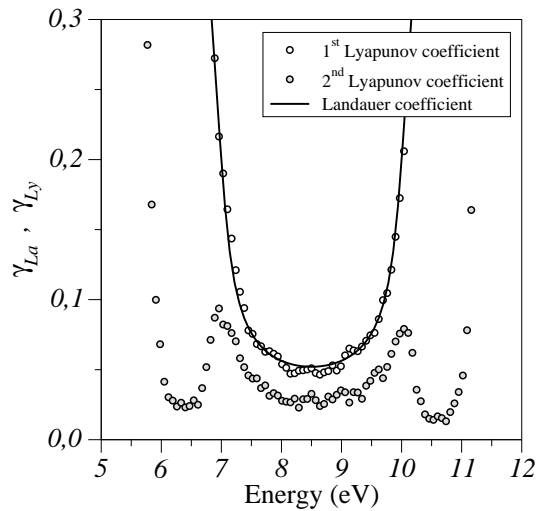}
\caption{$2/3$ of Landauer exponent $\gamma_\mathrm{La}$ (solid line) and the
largest Lyapunov exponent $\gamma_\mathrm{Ly}$ (white circles), as a function of
energy, for $t_\perp=0.5\,$eV and $t_\Vert=1.0\,$eV.  The second, smaller
Lyapunov exponent (grey circles) is also shown.}
\label{fig2}
\end{figure}

Figure~\ref{fig2} shows the Landauer and Lyapunov exponents for $N=4000$, when
$t_\perp=0.5\,$eV and $t_\Vert=1.0\,$eV. These values of the interstrand and
intrastrand hoppings are larger than those usually considered in the
literature,~\cite{Yan02,Klotsa05} but they were used by Caetano and
Schulz~\cite{Caetano05} to provide support to their claim about the extended
nature of the states. From Fig.~\ref{fig2} it becomes clear that neither the
largest Lyapunov exponents nor the Landauer one vanish over the whole energy
spectrum. Most important, its minimum value is size independent within the
numerical accuracy, suggesting the occurrence of truly localized states. Notice
that the minimum value of these exponents is always much larger than the
inverse of the number of base pairs ($1/N=0,00025$), indicating that DNA-pairing
can hardly explain long range charge transport at low temperature. From the
inverse of the minimum value of the second Lyapunov coefficient we can estimate
that the localization length is of the order of $80$ base pairs (i.e.\ roughly
eight turns of the double helix), therefore being much smaller than typical
sizes used in experiments.~\cite{Porath00}

\begin{figure}[ht]
\centering
\includegraphics[width=70mm,clip]{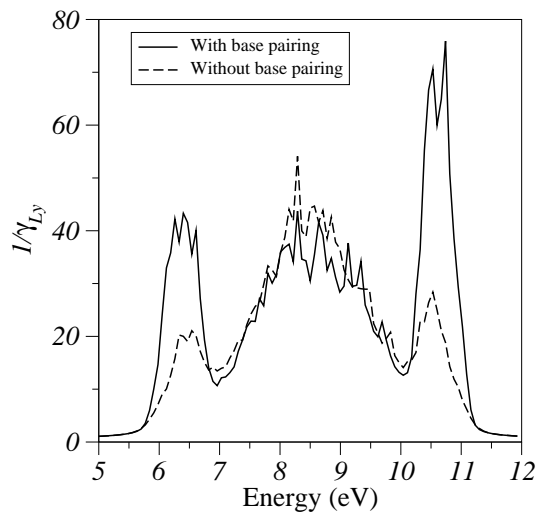}
\caption{Inverse of the second Lyapunov coefficient for $N=4000$
$t_\perp=0.5\,$eV and $t_\Vert=1.0\,$eV, when the base pairing is present (solid
line) and absent (dashed line).}
\label{fig3}
\end{figure}

To elucidate the effects of the base pairing on the localization length, we have
also considered the artificial case of ladder models without pairing. In that
case, both strands are completely random, allowing for a larger number of
possible pairs (e.g.\ AC or AA). Therefore, the system becomes much more
disordered and one could naively expect a dramatic decrease of the localization
length, as compared to the system with base pairing. Figure~\ref{fig3} indicates
that this is not the case. The inverse of the second Lyapunov exponent remains
almost unchanged over a large region of the energy spectrum, except close to the
two resonances at about $6.4\,$eV and $10.6\,$eV. At resonances the localization
length is reduced by a factor $2.5$ at most when the pairing constraint is
relaxed. In any event, resonances still appear so they cannot be associated to
base pairing.

We claimed that the spatial extend of states strongly depends on the hopping
parameters,~\cite{Sedrakyan06} and those considered in
Ref.~\onlinecite{Caetano05} seems to be larger as compared to those values
widely admitted in the literature.~\cite{Yan02,Klotsa05} Higher hoppings lead to
a \emph{less-effective\/} disorder and higher localization lengths are to be
expected. We have calculated the inverse of the second Lyapunov exponent for
more realistic values of the hopping parameter and checked that this claim is
indeed correct. For instance, for $t_\perp=0.05\,$eV and $t_\Vert=0.5\,$eV (see
Ref.~\onlinecite{Klotsa05}) the localization length at the center of the band is
reduced by a factor $5$ as compared to the case shown in Fig.~\ref{fig1}, while
an even larger decrease is noticed at resonances. Therefore, we come to the
conclusion that hopping is more important than base-pairing in this ladder
model.

\begin{figure}[ht]
\centering
\includegraphics[width=70mm,clip]{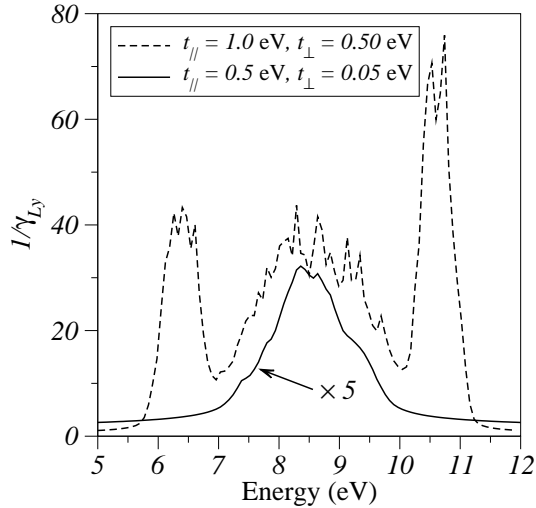}
\caption{Inverse of the second Lyapunov coefficient for $N=4000$ and two sets of
hopping parameters, indicated on the legend box. Notice the scaling factor
indicated on the lower curve.}
\label{fig4}
\end{figure}

\section{Conclusions} \label{conclusions}

We considered a ladder model of DNA for describing electronic transport in a
fully quantum-mechanical regime. In this model, a single orbital is assigned to
each base, and the sugar-phosphate channels are excluded. Therefore, it is
assumed that electronic conduction takes place through orbital overlap at the
bases. The sequence of bases of one of the strands is assumed to be totally
random, while the sequence of the other strand results from the base pairing
A--T and C--G.

We demonstrated analytically and numerically that due to the randomness of the
sequence, the states are always localized and a LDT cannot take place, contrary
to what claimed in Ref.~\onlinecite{Caetano05}. In particular, we observed that
base pairing has negligible effects on the localization length except close to
two resonant energies, located at about $6.4\,$eV and $10.6\,$eV for
$t_\perp=0.5\,$eV and $t_\Vert=1.0\,$eV. At these particular energies the
localization length is smaller when the pairing constraint is relaxed. Most
important, even in the case of base pairing, the larger localization length is
much smaller than typical sizes of samples used in various experiments.
Therefore, we come to the conclusion that base pairing alone is unable to
explain electronic transport at low temperature in DNA.

\acknowledgments

Work at Madrid was supported by MEC (Project MAT2003-01533). D.~S. and A.~S.
acknowledges INTAS grant 03-51-5460 for partial financial support.

\end{document}